\documentclass[12pt]{iopart}

\usepackage{iopams}
\usepackage[english]{babel}
\usepackage{graphicx, color, epsfig}
\usepackage{enumerate}
\usepackage{latexsym}
\usepackage{fancyhdr}
\usepackage{multirow}
\usepackage{cite}

\newcommand{\ds}{\displaystyle}
\newcommand{\vs}{\vspace}
\newcommand{\nt}{\nonumber}

\renewcommand{\a}{\textrm{\scriptsize a}}
\newcommand{\C}{\textrm{\scriptsize C}}
\newcommand{\CC}{\mathrm{C}}
\newcommand{\dd}{\mathrm{d}}
\newcommand{\ddd}{\mathbf{d}}
\newcommand{\E}{\mathbf{E}}
\renewcommand{\e}{\textrm{\scriptsize e}}
\newcommand{\ee}{\mathrm{e}}

\newcommand{\g}{\textrm{\scriptsize g}}

\newcommand{\I}{\mathrm{I}}
\newcommand{\J}{\mathrm{J}}
\renewcommand{\j}{\mathrm{j}}
\newcommand{\K}{\mathrm{K}}
\newcommand{\h}{\textrm{\scriptsize h}}
\newcommand{\kk}{\mathbf{k}}

\newcommand{\PP}{\mathrm{P}}
\newcommand{\rr}{\mathbf{r}}
\newcommand{\rrr}{\mathrm{r}}

\newcommand{\Y}{\mathrm{Y}}

\newcommand{\ldc}{[\![}
\newcommand{\rdc}{]\!]}
\newcommand{\pr}{\partial}

\renewcommand{\stackrel}[2]{\!\!\!\!
  \raisebox{-0.1cm}{$\begin{array}{c}
#1
\\[-0.15cm]
\textrm{\tiny $#2$}
  \end{array}$}\!\!\!\!}

\newcommand{\lnd}{~\!\!&}
\newcommand{\rnd}{&\!\!~}

\newcommand{\Si}{\textrm{Si}}

\begin{document}
\title[Stark Effect of Interactive Electron-hole pairs in Spherical Semiconductor QDs]{Stark Effect of Interactive Electron-hole pairs in Spherical Semiconductor Quantum Dots}

\author{B. Billaud$^{1,2}$\footnote{To whom correspondence should be addressed.}, M. Picco$^1$ and T.-T. Truong$^2$}
\ead{bbillaud@u-cergy.fr}
\
\address{$^1$Laboratoire de Physique Th\'eorique et Hautes Energies (LPTHE),
\\
CNRS UMR 7589, Universit\'e Pierre et Marie Curie (Paris VI),
\\
4, place Jussieu, 75252 Paris Cedex 05, France.}
\
\address{$^2$Laboratoire de Physique Th\'eorique et Mod\'elisation (LPTM),
\\
CNRS UMR 8089, Universit\'e de Cergy-Pontoise,
\\
2, avenue Adolphe Chauvin, 95302 Cergy-Pontoise Cedex, France.}
  \begin{abstract}
We present a theoretical approach, based on the effective mass approximation model, on the quantum-confinement Stark effects for spherical semiconducting quantum dots in the regime of strong confinement of interactive electron-hole pair and limiting weak electric field. The respective roles of Coulomb potential and polarization energy are investigated in details. Under reasonable physical assumptions, analytical calculations can be performed. They show that the Stark shift is a quadratic function of the electric field amplitude in this regime. The computed numerical values obtained from this approach are found to be in good agreement with experimental data over a significant domain of quantum dot sizes.
  \end{abstract}
\submitto{Journal of Physics: Condensed Matter}
\maketitle
\section{Introduction}
For about two decades, nano-structures, like quantum dots, quantum wires or quantum wells, are produced by diverse techniques such as etching, local inter-diffusion, particle suspension in dielectric media, or by self-assembly in matrices of a host material. They display many effects of standard atomic physics by restricting to a confined region of space the motion of one to a hundred embedded elementary charge carriers, which may be conduction band electrons, valence band holes, or excitons of the semiconducting host substrate. In contrast to atoms, two Quantum Dots (QDs) are never identical because phonons, surface effects and bulk disorder play a crucial role on their electronic properties. But, a QD may be considered as a giant artificial atom, which possesses an adjustable quantized energy spectrum, controlled by its size. Therefore, it enjoys prospects for an increasing range of future applications: {\it e.g.} as a semiconductor laser \cite{Kirkstaedter_1994} or as single-photon sources \cite{Yoshie_2004}, as {\it qubits} for quantum information processing \cite{Trauzettel_2007}, as single-electron transistors in micro-electronics \cite{Ishibashi_2003}, as artificial {\it fluorophores} for intra-operative detection of tumors, biological imaging or cell studies \cite{So_2006}.

Thanks to the progress of semiconductor growth technology during the early eighties, quantum size effects (QSE) showed up through optical properties of spherical semiconductor micro-crystals embedded in an insulating matrix \cite{Ekimov_1980, Golubkov_1981}. The characteristic blue-shift observed in optical spectra of such strongly quantum-confined systems emerges in a widening of semiconductor optical band gap, caused by the increasing confinement energy for decreasing QD size \cite{Brus_1984}. It has been also observed in a large variety of other confined micro-structures, like quantum ribbons or quantum disks \cite{Kash_1986}, quantum wires \cite{Temklin_1987} and quantum wells \cite{Vojak_1980}. A review of empirical and theoretical results on quantum confinement effects in low-dimensional semiconductor structures is given in \cite{Yoffe_2002}. The first theoretical attempt to describe semiconductor QDs has been elaborated upon a {\it particle-in-a-sphere} model, in the effective-mass approximation (EMA), which assumes parabolic valence and conduction bands \cite{Brus_1984, Efros_1982, Brus_1983, Brus_1986, Kayanuma_1988}. Both electron and hole behave as free particles, trapped in a spherical infinite potential well, and move with their common effective masses in the considered semiconductor. The electron-hole Coulomb interaction is included, and the excitonic contribution to the QD ground state energy is taken into account by Ritz' variational principle. Other EMA models have been built upon finite potential wells, improving agreement with experimental data for a significant range of QD sizes \cite{Kayanuma_1990, Kayanuma_1991, Nair_1987, Thoai_1990, Lo_1991}. In addition to spherical clusters, the case of cylindrical shaped micro-crystals has been carefully studied \cite{Kayanuma_1990, Potter_1988, Potter_1990, Le_Goff_1992}, as well as the case of quantum wires \cite{Brown_1986, Degani_1987}. More sophisticated models, which consider non-parabolic valence and/or conduction band(s), have been also developed \cite{Wang_1990, Nomura_1991, Xia_1989, Sercel_1990, Vahala_1990}. Modern approaches and how they can be applied to large structures and compared to experiments are discussed in the book \cite{Delerue} as well as in the recent review article \cite{Bester_2009}.

Among many important topics, it is the physics of atom-like behavior of QDs, which is nowadays most vigorously investigated in quantum confined systems, for potential technological applications. Of particular interest is the interaction with an ambient electromagnetic field, giving rise to the so-called quantum-confinement Stark effect (QCSE), which has been studied for example in $CdS_{0.12}Se_{0.88}$ \cite{Nomura_1990}, in $CdSe$ \cite{Bawendi_1997} or in $InGaAs$ micro-crystals \cite{Patane_2000}. It consists of an observable red-shift of an optical transition induced by the presence of a constant external electric field \cite{Yakimov_2003, Harutyunyan_2004, Ham_2005, Wei_2007}. In recent years, some works have also dealt with ac-electric field \cite{Bittencourt_2003, Rodriguez_2005}. Stark effect leads to an energy shift of the exciton photoluminescence as well as a corresponding enhancement of its recombination lifetime \cite{Polland_1985}. The electric field dependence of QCSE was first studied in $GaAs-AlGaAs$ multi-layers quantum wells \cite{Wood_1984}. Exciton energy shift peaks were experimentally observed and successfully compared to theoretical results obtained by a perturbation method introduced in \cite{Bastard_1983, Bastard_1984}, as the  applied electric field is perpendicular to the plane of the layer wells, within which the electron-hole Coulomb interaction is negligible. However, in spherical QDs, the Coulomb potential turns out to be more important, and cannot be discarded \cite{Nomura_1990, Nomura_1990b}.

Over the years, the spherical shape of QD has remained popular in the study of QCSE \cite{Chiba_1988, Ekimov_1990, Dissanayake_1995, Wen_1995, Chang_1998}. But, to the best of our knowledge, no simple comprehensive model, which describes Stark effects in spherical semiconductor nano-structures with analytic results, has been worked out. In this paper, we propose to use the EMA model for spherical micro-crystals, to establish analytically some criterions on the QD radius and on the electric field amplitude, and to understand why presently known results fail so far to correctly describe QCSE for a wide range of QD radius. To this end, in Section {\bf \ref{sec_II}}, we shall introduce the electric field free model first, and recall some of its general properties. The next two Sections {\bf \ref{sec_III}} and {\bf \ref{sec_IV}} are devoted to the analysis of Stark effects in spherical semiconductor QDs, first with the inclusion of electron-hole Coulomb potential and second with an additional polarization energy. In the concluding section, we summarize our main results and indicate possible future research perspectives.
\section{EMA Quantum Dot model} \label{sec_II}
A standard EMA model with infinite spherical confining potential well, without electron-hole spin coupling and external magnetic field, allows to perform analytically most of the calculations on spherical semiconductor nano-structures interacting with a fixed external electric field. There exists other models with parabolic confinement \cite{Keller_1995, Lozovik_2003} or parabolic potential superimposed to an infinite potential well \cite{Sundqvist_2002}, but the concept of a QD size is then not so well defined. As Stark effect in semiconducting micro-crystallites manifests itself through an energy shift of the electron-hole total energy levels, we have to deal mainly with energy eigenvalue differences of a Hamiltonian. However, two different energies have to be computed within the same theoretical QD model. Even if this model does not fully describe the QD behavior in the absence of electric field, particularly for small QD radii, it can be still used, since it gives rather satisfactory theoretical predictions on Stark effect. The overestimation made for the electric field free electron-hole pair energy levels should also appear in the interaction of the electric field with electron-hole pairs \cite{Thoai_1990}. Then, despite intrinsic limitations, an approximate value of the Stark shift can be obtained under some consistency conditions.
\subsection{Consistency conditions} \label{subsec_II_A}
Since most synthesized nano-crystallites possess an {\it aspect ratio} (defined as the ratio between the longest and shortest axes of the QD) smaller than 1.1, even if higher aspect ratio micro-crystals would exist, the hypothesis of a QD with spherical symmetry appears often as quite reasonable.

Realistically, the effective potential at the QD surface is finite, and has a standard amplitude from 1 to 3eV \cite{Thoai_1990}. This value justifies already the use of an infinite confining potential well, as mentioned earlier, because it is generally quite large as compared to typical electron and hole confinement energies usually involved, which increase as $\propto R^{-2}$ for decreasing QD radius $R$ \cite{Brus_1983}. Therefore, the tunnel conductivity through the QD boundary is vanishingly small, except of course for very small QDs. Futhermore, the infinite potential well approximation implies that charge carriers inside the cluster are insensitive to its outside surroundings, particularly to any externally applied field, as far as considerations on QCSE are concerned. Although the surrounding effects may be sufficiently small to be neglected, the presence of a large external field can significantly modify the inside behavior of the micro-crystallites. Thus, the electric field amplitude outside the QD should not then exceed a threshold, fixed {\it ad hoc} by the height of the real confining potential step. This constraint should be referred to as the usual weak electric field limit. An inequality, which analytically expresses its validity by linking the electric field amplitude to other physical parameters of the problem, is to be determined later. It allows to evaluate an approximate value for the maximal electric field amplitude to apply on the QD, while respecting the weak field limit.

Lastly, for small nano-crystals of typical sizes of less than a hundred lattice spacings, there exist {\it magic numbers} for which clusters remain stable: {\it e.g.} crystalline silicon only stay coherent as clusters of $Si_{12}, Si_{33}, Si_{39}$ and $Si_{45}$, if they contain less than 60 silicon atoms \cite{Pan_1994}. Their band structure are so deformed such that it becomes impossible to use the parabolic shape of conduction and valence bands, required by EMA models. However, if there is no potential well --- {\it i.e.} if there is no semiconductor micro-crystal embedded in the surrounding insulating matrix ---, no electrons should be excited, and no holes should appear. Stark effects must then vanish in small QDs for any electric field. In the weak field limit, it has been shown that, in semiconducting rectangular quantum boxes, the Stark shift of the confined exciton ground state presents three contributions, each of them going as the fourth power of an edge length \cite{Wei_2007}. Hence, for a spherical potential well of radius $R$, the Stark shift is expected to scale as $\propto R^4$.
\subsection{General considerations} \label{subsec_II_B}
We work with the infinite confining potential well $\ds
V(\rr)=V(r)=\!
    \left\{
      \begin{array}{ccc}
0 & \textrm{if} & 0\leq r\leq R
\\
\infty & \textrm{if} & r>R
      \end{array}
    \right.\!\!$, written in spherical coordinates $(r,\theta,\varphi)$. According to the EMA model, the total electron-hole pair Hamiltonian operator $H_0$ reads, in units where $\hbar=1$
  \begin{eqnarray}
H_0\lnd=\rnd H_\e+H_\h+V_\C(\rr_{\e\h})=-\frac{\nabla^2_{\!\e}}{2m_\e^*}-\frac{\nabla^2_{\!\h}}{2m_\h^*}+V(\rr_\e)+V(\rr_\h)-\frac{e^2}{\kappa r_{\e\h}},~~~ \label{H_0}
  \end{eqnarray}
where $\kappa=4\pi\varepsilon$, $\varepsilon$ denotes the semiconductor dielectric constant, $r_{\e\h}=|\rr_{\e\h}|=|\rr_\e-\rr_\h|$ the electron-hole relative distance, $m_{\e,\h}^*$ the effective mass and $H_{\e,\h}$ the confinement Hamiltonian respectively of the electron and of the hole and $V_\C(\rr_{\e\h})$ the electron-hole Coulomb interaction. Without loss of generality, the semiconductor energy band gap $E_\g$ is set equal to be zero for convenience.

In the absence of Coulomb potential, electron and hole are treated as decoupled particles, the QD wave function should be then factorized into separable electronic and hole parts $\Psi(\rr_\e,\rr_\h)=\psi(\rr_\e)\psi(\rr_\h)$. The orthonormal eigenfunctions $\psi_{lnm}(\rr)$ are labeled by three quantum numbers $l\!\in\!\mathbb{N}$, $n\!\in\!\mathbb N^*=\mathbb N\smallsetminus\{0\}$ and $m\!\in\!\ldc-l,l\rdc$. If $0\leq r\leq R$,
  \begin{equation}
\psi_{lnm}(\rr)=\psi_{lnm}(r,\theta,\varphi)=\sqrt{\frac2{R^3}}\frac{\Y^m_l(\theta,\varphi)}{\j_{l+1}(k_{ln})}\j_l\!\!\left(\frac{k_{ln}}Rr\!\right)\!,
  \end{equation}
where $\Y^m_l(\theta,\varphi)$ is the spherical harmonic of orbital quantum number $l$ and azimuthal quantum number $m$, $\j_l(x)$ the spherical Bessel function of the first kind of index $l$ and of variable $x$, and $\left\{k_{ln}\right\}_{ln}$ the wave numbers set, defined as the $n^{\textrm{\scriptsize th}}$ non-zero root of the spherical Bessel function $\j_l(x)$ thanks to the continuity condition at $r=R$ --- the presence of an infinite potential imposes that $\psi_{lnm}(\rr)=0$, if $r>R$ \cite{Efros_1982}. The respective energy eigenvalues for electron and hole are expressed in terms of $\{k_{ln}\}_{ln}$ as
  $$
E^{\e,\h}_{ln}=\frac{k_{ln}^2}{2m^*_{\e,\h}R^2}.
  $$
The continuum density of states of the semiconductor bulk should show atomic-like discrete spectrum with increasing energy separation as the radius decreases.

Because of the explicit micro-crystallites spherical symmetry breakdown in the presence of a Coulomb potential, the exact determination of eigenfunctions and energy eigenvalues for Eq. \eref{H_0} is arduous. Treating the interplay of the Coulomb interaction, scaling as $\propto\! R^{-1}$, and the quantum confinement, scaling as $\propto R^{-2}$, constitutes the common approach to this problem. To handle these competing contributions, two working regimes are singled out, according to the ratio of the QD radius $R$ to the Bohr radius of the bulk Mott-Wannier exciton $\ds a^*=\frac{\kappa}{e^2\mu}$, $\mu$ being the reduced mass of the exciton \cite{Kayanuma_1988}. In the strong confinement regime, corresponding to a size $R\leq2a^*$, the potential well strongly affects the relative electron-hole motion, and {\it exciton} states consist of uncorrelated electron and hole states. The weak confinement regime, valid for a size $R\geq4a^*$, leaves the electron-hole relative motion and its binding energy unchanged. The exciton character of a quasi-particule of total mass $M=m_\e^*+m_\h^*$ is conserved. As its center-of-mass motion remains confined, it should be quantized \cite{Kayanuma_1988}.

Even if we focus on Stark effect in the strong confinement regime, we shall briefly present the consequences of the previous single EMA model in both strong and weak confinement regimes. Despite its simplicity, this model seems to be able to apprehend correctly the QCSE, at least for a range of sufficiently small QD sizes, and to yield numerical values, which agree with experimental results.
\subsection{Considerations on strong and weak confinement regimes}
\subsubsection{Strong confinement regime} \label{subsubsec_II_C_1}
\

\noindent In this regime, the Coulomb potential is treated as a perturbation with respect to the infinite confinement potential well in a variational procedure, which shall be extended to the case of an applied electric field. The ground state energy of the exciton should be evaluated with the following trial wave function
  \begin{equation}
\phi(\rr_\e,\rr_\h)=\psi_{010}(\rr_\e)\psi_{010}(\rr_\h)\phi_{\textrm{\scriptsize rel}}(\rr_{\e\h}), \label{phi}
  \end{equation}
with $\phi_{\textrm{\scriptsize rel}}(\rr_{\e\h})=\phi_{\textrm{\scriptsize rel}}(r_{\e\h})=\ee^{-\frac\sigma2r_{\textrm{\tiny eh}}}$, where $\sigma$ is the variational parameter, $r_{\e,\h}=|\rr_{\e,\h}|$ and
  $$
\ds\psi_{010}(\rr_{\e,\h})\!=\!\psi_{010}(r_{\e,\h})\!=\!\frac{\sin\!\left(\frac\pi R r_{\e,\h}\right)}{r_{\e,\h}\sqrt{2\pi R}}.
  $$
The variational wave function of Eq. \eref{phi} implies that both electron and hole should occupy primarily their respective ground state in the confining infinite potential well, as described by the product $\psi_{010}(\rr_\e)\psi_{010}(\rr_\h)$. It should also exhibit, {\it via} the function $\phi_{\textrm{\scriptsize rel}}(\rr_{\e\h})$ of the relative coordinates $\rr_{\e\h}$, an exciton bound state behavior, analogous to the ground state of an hydrogen-like atom with appropriate mass $\mu$ and Bohr radius $a^*$, up to a normalization factor, especially if $\sigma^{-1}\propto a^*$.

Despite the breakdown of translational invariance of the Coulomb interaction by the spherical confining potential, Fourier transform formalism in relative electron-hole coordinates allows to establish integral representations for quantities such as the square of the norm $\langle\phi|\phi\rangle$ of the trial function $\phi(\rr_\e,\rr_\h)$ or the corresponding Coulomb potential diagonal matrix element $\langle\phi|V_\C(r_{\e\h})|\phi\rangle$
  \begin{equation}
\fl\left\{
    \begin{array}{rcl}
\ds\langle\phi|\phi\rangle\!\!\!\!\lnd=\rnd\!\!\!\!\ds-\frac8{R^2}\frac\pr{\pr\sigma}\!\left[\frac1\sigma\!\int\!\!\!\!\int_{\mathcal D}\!\frac{\dd x}x\frac{\dd y}y\sin^2(\pi x)\sin^2(\pi y)\sinh(\sigma Rx)\ee^{-\sigma Ry}\right]\!,
\vs{.2cm}
\\
\ds\langle\phi|V_\C(r_{\e\h})|\phi\rangle\!\!\!\!\lnd=\rnd\!\!\!\!\ds-\frac{e^2}{\kappa R}\frac8\sigma\!\int\!\!\!\!\int_{\mathcal D}\!\frac{\dd x}x\frac{\dd y}y\sin^2(\pi x)\sin^2(\pi y)\sinh(\sigma Rx)\ee^{-\sigma Ry},
    \end{array}
  \right. \label{phi|phi}
  \end{equation}
where $\mathcal D=\{(x,y)\in\mathbb R^2~/~0\leq x\leq y\leq1\}$. A Taylor expansion of expressions Eqs. \eref{phi|phi} with respect to the dimensionless parameter $\sigma R$ near zero yields
  $$
    \left\{
      \begin{array}{rcl}
\langle\phi|\phi\rangle\!\!\!\!\lnd=\rnd\!\!\!\! 1-B\sigma R+\Or(\sigma^2R^2),
\vs{.2cm}
\\
\langle\phi|V_\C(r_{\e\h})|\phi\rangle\!\!\!\!\lnd=\rnd\!\!\!\!\ds-\frac{e^2}{\kappa R} \left\{A-\sigma R+\Or(\sigma^2R^2)\right\}\!.
      \end{array}
    \right.
  $$
Thus, an expression of the mean value of the total Hamiltonian $H_0$ in the strong confinement regime in terms of a dimensionless variational parameter $\sigma'$, defined by $\sigma'=\sigma a^*$, and of the binding exciton Rydberg energy $\ds E^*=\frac1{2\mu a^{*2}}$ can be obtained as
  $$
\frac{\langle\phi|H_0|\phi\rangle}{\langle\phi|\phi\rangle}=E_{\e\h}-A\frac{e^2}{\kappa R}-2B'E^*\sigma'+\frac{E^*}4\sigma'^2+\dots
  $$
where the correction terms ``~\dots~'' go to zero as soon as $\ds\frac R{a^*}$ goes to zero, and where $\ds E_{\e\h}=E^\e_{01}+E^\h_{01}$ is the electron-hole pair ground state confinement energy \footnote{\ref{appendix_B} compiles a register of all constants, which appear in this paper.}. The variational parameter $\sigma'$ is determined to be $\sigma'_0=4B'\approx0.9956$, to minimize the value of the electron-hole energy
  $$
E^{\textrm{\scriptsize strong}}_{\e\h}= E_{\e\h}-A\frac{e^2}{\kappa R}-4B'^2E^*,
  $$
This formula has been already analytically obtained with trial functions showing the same global form as $\phi(\rr_\e,\rr_\h)$, but with an interactive part equal to $\widetilde\phi_{\textrm{\scriptsize rel}}(\rr_{\e\h})=(1-\frac\sigma2r_{\e\h})$, instead of $\phi_{\textrm{\scriptsize rel}}(\rr_{\e\h})$ \cite{Kayanuma_1988, Schmidt_1986}. It is obvious that $\widetilde\phi_{\textrm{\scriptsize rel}}(\rr_{\e\h})$ consists of the two first terms of the Taylor expansion of $\phi_{\textrm{\scriptsize rel}}(\rr_{\e\h})$, in the limit $\frac\sigma2r_{\e\h}\leq\sigma R\ll1$. Because of the infinite confining potential well assumption, the total excitonic energy is overestimated in comparison to experimental data for small QDs. A successful method to subtract off this over-estimation consists in adopting a model in which a confining finite potential step of experimentally acceptable height is restored \cite{Thoai_1990}.
\subsubsection{Weak confinement regime}
\

\noindent In this regime, electron-hole pair states consist of exciton bound states. The Coulomb potential and the kinetic energy in the electron-hole relative coordinates are of the same order of magnitude because the QD size allows a partial restoration of the long range Coulomb interaction between the charged carriers inside the QD. Then, the essential contribution to the ground state energy of the exciton is $-E^*$, the ground state energy of a hydrogen-like atom mass $\mu$. Furthermore, the total translational motion of the exciton, thought as a quasi-particle of total mass $M$, should be restored and contributes to the exciton total energy by $\ds\frac{\pi^2}{2MR^2}$, the typical kinetic energy term of a free particle confined in a space region of size $R$. As a first approximation, the ground state energy of the exciton trapped inside the QD is then the sum of these two energetic contributions, but this is not totally satisfactory. To improve phenomenologically its accuracy in regard to numerical simulations, a monotonic increasing function $\eta(\lambda)$ of the effective masses ratio $\ds\lambda=\frac{m_\h^*}{m_\e^*}$ has been introduced in \cite{Kayanuma_1988}, and has been inserted into $E^{\textrm{\scriptsize weak}}_{\e\h}$ as follows
  $$
E^{\textrm{\scriptsize weak}}_{\e\h}=-E^*+\frac{\pi^2}{2M(R-\eta(\lambda)a^*)^2}.
  $$
The QD size {\it renormalization} term $\eta(\lambda)a^*$ is a dead layer \cite{Hopfield_1963}. Although it could be successfully described as a quasi-particle, the exciton is not itself an indivisible particle. Its center-of-mass, whose motion is quantized, could not reach the infinite potential well surface unless the electron-hole relative motion undergoes a strong deformation. The picture of a point-like exciton should be dropped in this region of space. The exciton should be preferentially thought as a rigid sphere of radius $\eta(\lambda)a^*$, where $\eta(\lambda)$ is numerically determined to get a better fit of experimental results \cite{Kayanuma_1988}.
\section{QD Stark effect without polarization energy} \label{sec_III}
The diagonalization of the Hamiltonian of an electron-hole pair trapped in an infinite potential well under the influence of an external constant electric field is, in principle, an exactly solvable problem. Even if the presence of a constant electric field explicitly breaks both spherical QD symmetry and electron-hole Coulomb potential translation invariance, in a spherical QD, we shall consider an applied electric field $\E_\a$ along the direction $z$ of a three-dimensional cartesian coordinates system with its origin located at the QD center. This is not the cases of quantum wells \cite{Bastard_1983, Bastard_1984}, quantum wires \cite{Spector_2005}, quantum disks \cite{Ham_2005}, rectangular \cite{Wei_2007}, cubical \cite{Spector_2007} or confined by parabolic potential \cite{Pacheco_1997} quantum boxes, in which the electric field direction plays a significant role. As the inside semiconducting QD dielectric constant $\varepsilon$ is larger than the outside insulating matrix dielectric constant $\varepsilon'$, the electric field $\E_\dd$ inside the QD is different from $\E_\a$. It is given by $\E_\dd=\ds\frac{\E_\a}{(1-g)+g\varepsilon_\rrr}$, where $g$ is a geometrical depolarization factor, which equals $\ds\frac13$ for a sphere, and $\varepsilon_\rrr=\ds\frac\varepsilon{\varepsilon'}$ is defined as the relative dielectric constant \cite{Magid}. The dielectric constant difference also implies the existence of a polarization energy term $P(\rr_\e,\rr_\h)$, introduced in \cite{Brus_1984}, which shall be neglected in this section, but taken into account in the next one. This hypothesis allows to study in more details its relative role {\it vs.} the Coulomb potential, because they both scale as $\propto R^{-1}$.

Let us define the electron and the hole (of respective electric charge $\mp e$) interaction Hamiltonian with the electric field $\E_\dd$, in spherical coordinates, as
  \begin{eqnarray}
W_{\e,\h}(\rr_{\e,\h})=\pm e\E_\dd\cdot\rr_{\e,\h}=\pm eE_\dd r_{\e,\h}\cos\theta_{\e,\h},
  \end{eqnarray}
where $E_\dd=|\E_\dd|$ is the electric field amplitude inside the micro-crystal. As the function $\phi(\rr_\e,\rr_\h)$ does not provide any further contribution to the excitonic energy in the presence of the electric field, {\it i.e.} $\langle\phi|W_\e(\rr_\e)|\phi\rangle=-\langle\phi|W_\h(\rr_\h)|\phi\rangle$, an appropriate form for the trial wave function should present some other dependence on the electron and hole space coordinates to be determined later.
\subsection{Justification of the variational trial wave function form} \label{subsec_III_A}
To apprehend the effect of the induced electric polarization, we follow a reasoning made in \cite{Bastard_1983}, and study the interaction between the charge carriers with the ambient electric field, neglecting the Coulomb potential. To this end, we define the individual Hamiltonian $H'_{\e,\h}$ of a confined electron or of a confined hole interacting with $\E_\dd$ as
  $$
H'_{\e,\h}= H_{\e,\h}+W_{\e,\h}(\rr_{\e,\h}).
  $$
As mentioned as a consistency condition in Subsection {\it \ref{subsec_II_A}}, we can assume that the electric field amplitude is sufficiently small so as to consider the Hamiltonian interactive part $W_{\e,\h}(\rr_{\e,\h})$ as a perturbation to the confined Hamiltonian $H_{\e,\h}$. In this weak field limit, the absolute typical interaction energy of the electron or of the hole under the electric field influence $E_{\textrm{\scriptsize ele}}=eE_\dd R$ should be treated as a perturbation compared to their typical confinement energy $E_{\e,\h}=E^{\e,\h}_{01}$, {\it i.e.} $E_{\textrm{\scriptsize ele}}\ll E_{\e,\h}$. In real atoms as well as in quantum wells \cite{Bastard_1984}, in quantum wires \cite{Spector_2005} and quantum boxes \cite{Spector_2007}, Stark shifted levels show a typical quadratic dependence on the electric field amplitude, a similar behavior is expected here. In the following, to justify the form of the variational wave function leading to the QCSE, we investigate perturbative results on decoupled confined electron and hole, interacting individually with an external electric field, by performing first, a standard second-order stationary perturbation theory, and second, a variational procedure.
%
%
%

A second-order perturbation computation of the Stark shift undergone by the electron or by the hole ground state, yields
  $$
\Delta E_{\e,\h}^{\textrm{\scriptsize Stark pert}}=-\Gamma_{\textrm{\scriptsize pert}}m_{\e,\h}^*e^2E_\dd^2R^4+\Or(E_\dd^3),
  $$
where the constant $\ds\Gamma_{\textrm{\scriptsize pert}}=\frac{32}3\pi^2\sum_{n\geq1}\frac{k_{1n}^2}{(k_{1n}^2-\pi^2)^5}\approx0.01817$.
%
%
%

In order to account for the electric field direction along the $z$-axis in the variational principle, the trial wave function should show a deformation away from the spherical shape, which squeezes or stretches the electron or the hole probability density along this particular direction. The variational trial function is chosen of the form $\Phi_{\e,\h}(\rr_{\e,\h})=\psi_{010}(\rr_{\e,\h})\varphi_{\e,\h}(\rr_{\e,\h})$, where $\varphi_{\e,\h}(\rr_{\e,\h})=\ee^{\mp\frac{\sigma_{\textrm{\tiny e,h}}}2r_{\textrm{\tiny e,h}}\cos\theta_{\textrm{\tiny e,h}}}$. The variational parameters $\sigma_{\e,\h}$ have the dimension of an inverse length so that, in the weak field approximation, we can assume that $\sigma_{\e,\h}R\ll1$. The difficulty in this problem is in the calculation of the square of the norm of the trial function $\Phi_{\e,\h}(\rr_{\e,\h})$. However, it admits an integral representation, on which a Taylor expansion in the neighborhood of the dimensionless parameter $\sigma_{\e,\h}R=0$ can be performed
  \begin{eqnarray*}
\langle\Phi_{\e,\h}|\Phi_{\e,\h}\rangle\lnd=\rnd\frac2{\sigma_{\e,\h}R}\int_0^1\!\frac{\dd x}x\sin^2(\pi x)\sinh(\sigma_{\e,\h}Rx)
\\
\lnd=\rnd1+\frac C6\sigma^2_{\e,\h}R^2+\Or\!\left(\sigma^4_{\e,\h}R^4\right)\!.
  \end{eqnarray*}
The mean value of the confinement Hamiltonian $H_{\e,\h}$ in the quantum state defined by the trial function $\Phi_{\e,\h}(\rr_{\e,\h})$ is exactly determined as
  $$
\frac{\langle\Phi_{\e,\h}|H_{\e,\h}|\Phi_{\e,\h}\rangle}{\langle\Phi_{\e,\h}|\Phi_{\e,\h}\rangle}=E_{\e,\h}+\frac{\sigma^2_{\e,\h}}{8m_{\e,\h}^*},
  $$
and the mean value of the interaction Hamiltonian $W_{\e,\h}(\rr_{\e,\h})$ as
  $$
\frac{\langle\Phi_{\e,\h}|W_{\e,\h}|\Phi_{\e,\h}\rangle}{\langle\Phi_{\e,\h}|\Phi_{\e,\h}\rangle}=-e\sigma_{\e,\h}E_\dd R^2\!\left\{\frac C3+\Or\!\left(\sigma^2_{\e,\h}R^2\right)\right\}\!.
  $$
Then, the total Hamiltonian $H'_{\e,\h}$ mean value, up to the second order in $\sigma_{\e,\h}R$, is
  $$
\frac{\langle\Phi_{\e,\h}|H'_{\e,\h}|\Phi_{\e,\h}\rangle}{\langle\Phi_{\e,\h}|\Phi_{\e,\h}\rangle}=
E_{\e,\h}+\frac{\sigma^2_{\e,\h}}{8m_{\e,\h}^*}-\frac C3eE_\dd
R^2\sigma_{\e,\h}+\dots
  $$
A choice of the variational parameter $\ds\sigma^0_{\e,\h}=\frac{4C}3m_{\e,\h}^*eE_\dd R^2$ gives the ground state energy $E'_{\e,\h}$ of the confined charge carriers in interaction with the electric field, and the Stark shift by subtracting their respective ground state energy, as follows
  $$
\Delta E^{\e,\h}_{\textrm{\scriptsize Stark var}}=E'_{\e,\h}-E_{\e,\h}=-\Gamma_{\textrm{\scriptsize var}}m_{\e,\h}^*e^2E_\dd^2R^4,
  $$
where $\ds \Gamma_{\textrm{\scriptsize var}}=\frac{2C^2}9\approx0.01776$.

These two Stark shift expressions present the same dependence on physical parameters: they both scale as $\propto m_{\e,\h}^*e^2E_\dd^2R^4$. The Stark shift contribution is clearly a second order term in the dimensionless parameter $\ds\frac{E_{\textrm{\scriptsize ele}}}{E_{\e,\h}}\ll1$, with respect to the electron or the hole confinement energy $E_{\e,\h}$. 
The difference between the previous methods is quantifiable by evaluating the relative error between the values of the proportionality constants $\Gamma_{\textrm{\scriptsize pert}}$ and $\Gamma_{\textrm{\scriptsize var}}$, which is $\approx2\%$. This small relative error supports the validity of the new trial wave function $\Phi(\rr_\e,\rr_\h)$ in the presence of the electric field, defined as
  \begin{equation}
\Phi(\rr_\e,\rr_\h)=\phi(\rr_\e,\rr_\h)\varphi_\e(\rr_\e)\varphi_\h(\rr_\h). \label{Phi}
  \end{equation}
Since, the function $\Phi(\rr_\e,\rr_\h)$ has the part $\phi(\rr_\e,\rr_\h)$, describing the electron-hole Coulomb interaction both occupying the ground state of their respective confinement Hamiltonian, and the electric field interactive part $\varphi_\e(\rr_\e)\varphi_\h(\rr_\h)$, liable for the individual electron and hole behaviors in the ambient electric field $\E_\dd$.
\subsection{General results on Stark effect in semiconductor Quantum Dots} \label{subsec_III_B}
As already mentioned, we add the interaction Hamiltonians $W_{\e,\h}(\rr_{\e,\h})$ between the charge carriers and the electric field to the QD model Hamiltonian $H_0$, introduced in Section {\bf \ref{sec_II}}, in order to apprehend QCSE in spherical semiconductor micro-crystals
  \begin{equation}
H=H_0+W_\e(\rr_\e)+W_\h(\rr_\h).
  \end{equation}
In the weak field limit, the variational procedure is to be applied using the trial function $\Phi(\rr_\e,\rr_\h)$, introduced in Eq. \eref{Phi}. To this end, we use a reasoning similar to that of Subsection {\it \ref{subsec_II_B}}, {\it i.e.} the Fourier transform formalism in the relative coordinates can be used once again quite advantageously. This formalism leads to integral representation of the square of the norm of the trial function $\Phi(\rr_\e,\rr_\h)$ and of the mean value of the Coulomb interaction matrix element in the corresponding quantum state
  \begin{equation}
\fl\left\{
  \begin{array}{rcl}
\langle\Phi|\Phi\rangle\!\!\!\!\lnd=\rnd\!\!\!\!\ds-\frac2{R^2}\frac\pr{\pr\sigma}\!\left[\frac1\sigma\!\int_{-1}^1\!\dd\xi\!\int\!\!\!\!\int_{\mathcal D}\!\frac{\dd x}x\frac{\dd y}y\sin^2(\pi x)\sin^2(\pi y)\right.
\vs{.2cm}
\\
\lnd\rnd\!\!\!\!\!\!\!\!\!\!\!\!\!\!\!\!\!\!\!\!\!\!\!\!\left.\ds\times\!\left\{\frac{\sinh\!\left(\rho_\e(\xi)\sigma Rx\right)}{\rho_\e(\xi)}\frac{\ee^{-\rho_\textrm{\tiny h}(\xi)\sigma Ry}}{\rho_\h(\xi)}+\frac{\sinh\!\left(\rho_\h(\xi)\sigma Rx\right)}{\rho_\h(\xi)}\frac{\ee^{-\rho_\textrm{\tiny e}(\xi)\sigma Ry}}{\rho_\e(\xi)}\right\}\!\right]\!\!,
\vs{.2cm}
\\
\langle\Phi|V_\C(r_{\e\h})|\Phi\rangle\!\!\!\!\lnd=\rnd\!\!\!\!\ds-\frac{e^2}{\kappa R}\frac2{\sigma R}\int_{-1}^1\!\dd\xi\!\int\!\!\!\!\int_{\mathcal D}\!\frac{\dd x}x\frac{\dd y}y\sin^2(\pi x)\sin^2(\pi y)
\vs{.2cm}
\\
\lnd\rnd\!\!\!\!\!\!\!\!\!\!\!\!\!\!\!\!\!\!\!\!\!\!\!\!\ds\times\!\left\{\frac{\sinh\!\left(\rho_\e(\xi)\sigma Rx\right)}{\rho_\e(\xi)}\frac{\ee^{-\rho_\textrm{\tiny h}(\xi)\sigma Ry}}{\rho_\h(\xi)}+\frac{\sinh\!\left(\rho_\h(\xi)\sigma Rx\right)}{\rho_\h(\xi)}\frac{\ee^{-\rho_\textrm{\tiny e}(\xi)\sigma Ry}}{\rho_\e(\xi)}\right\}\!,
  \end{array} \label{Phi|Phi}
\right.
  \end{equation}
where $\ds\rho_{\e,\h}(\xi)=\sqrt{1-2\frac{\sigma_{\e,\h}}\sigma\xi+\frac{\sigma^2_{\e,\h}}{\sigma^2}}$, for $-1\leq\xi\leq1$.

Direct calculations prove that Eqs. \eref{Phi|Phi} are valid if and only if the variational parameters $\sigma$ and $\sigma_{\e,\h}$ satisfy the inequality
  \begin{equation}
\ds0\leq\ee\cdot\sigma_{\e,\h}<\sigma, \label{consistency_condition}
  \end{equation}
where $\ee=\exp(1)$. Hence, Eq. \eref{consistency_condition} is a consistency condition, which analytically determines the range of acceptable electric field amplitudes. First, following the variational results for the interaction between the electric field with the electron or with the hole, we expect that $\sigma_{\e,\h}\propto m_{\e,\h}^*eE_\dd R^2$. Second, following the variational results on the electric field free interactive electron-hole pair through the Coulomb potential, we also expect that $\sigma\propto a^{*-1}$. Then, after trivial rearrangements, we remark that $\ds\frac{E_{\textrm{\scriptsize ele}}}{E_{\e,\h}}\propto\frac{\sigma_{\e,\h}}\sigma\frac R{a^*}$, {\it i.e.} the weak field limit should remain valid if $\ds\frac{E_{\textrm{\scriptsize ele}}}{E_{\e,\h}}$ does not exceed, up to a dimensionless proportionality factor to be given later, the order of magnitude of $\ds\frac R{a^*}$, the ratio which characterizes the strong confinement regime. The charge carrier energy, in the presence of the electric field, should be at most of the same order of magnitude of a first term correction to their confinement energy in the strong confinement regime, which corresponds to the absolute value of the typical electron-hole Coulomb interaction energy, because they both scale as $\propto R^{-1}$.

In the limit of vanishing electric field, {\it i.e.} in the limit $\ds\frac{\sigma_{\e,\h}}\sigma\rightarrow0$, the expressions in Eqs. \eref{Phi|Phi} allow to retrieve the expressions for the square of the norm and for the Coulomb potential mean value without electric field expressed by Eq. \eref{phi|phi}. Moreover, in the weak field limit, based on the decoupled electron-hole point of view presented in Subsection {\it \ref{subsec_III_A}}, we expect that the Stark shift for the coupled electron-hole system should scale as $\ds\propto (m_\e^*+m_\h^*)e^2E_\dd^2R^4\propto E_{\e\h}\frac{R^2}{a^{*2}}$. Then, to get at least the lowest order contribution to this Stark shift, it is necessary to perform a Taylor expansion of the total Hamiltonian $H$ mean value up to the second order in the variational parameters. However, as we shall see in the following subsection, this first contribution is not sufficiently accurate to fit experimental data, because it does not account for the electron-hole coupling through the Coulomb interaction. This is the reason why we shall carry on the expansion up to the third order, since we will also get the first correction in $\ds\frac R{a^*}$ to the Stark shift, which expresses the presence of the Coulomb potential in the strong confinement regime.

Finally, we obtain the interaction Hamiltonian $W_{\e,\h}(\rr_{\e,\h})$ mean value from the square of the norm of the wave function $\Phi(\rr_\e,\rr_\h)$ by taking its logarithmic derivative with respect to the variational parameters $\sigma_{\e,\h}$
  \begin{equation}
\frac{\langle\Phi|W_{\e,\h}(\rr_{\e,\h})|\Phi\rangle}{\langle\Phi|\Phi\rangle}=-eE_\dd\frac\pr{\pr\sigma_{\e,\h}}\log\langle\Phi|\Phi\rangle.
  \end{equation}
As we have build the trial function $\Phi(\rr_\e,\rr_\h)$, so that it possesses the properties of the functions $\phi(\rr_\e,\rr_\h)$ and $\Phi_{\e,\h}(\rr_{\e,\h})$, the exact mean value of the electric field free Hamiltonian $H_0$ should lead to
  \begin{equation}
\frac{\langle\Phi|H_0|\Phi\rangle}{\langle\Phi|\Phi\rangle}=E_{\e\h}+\frac{\sigma^2}{8\mu}+\frac{\sigma^2_\e}{8m^*_\e}+\frac{\sigma^2_\h}{8m^*_\h}, \label{Phi|H_0|Phi}
  \end{equation}
where the electron-hole pair ground state confinement energy $E_{\e\h}$ is provided by the function $\psi_{010}(\rr_\e)\psi_{010}(\rr_\h)$, and the contributions $\ds\frac{\sigma^2}{8\mu}$ and $\ds\frac{\sigma^2_{\e,\h}}{8m^*_{\e,\h}}$ to the total kinetic energy are respectively due to the Coulomb potential and to the interaction between the charge carriers and the electric field. However, in addition to these four expected terms, direct calculations exhibit a further contribution to the mean value of $H_0$ of the form
  $$
K(\sigma,\sigma_\e,\sigma_\h)=\frac\sigma4\!\left\{\frac\pr{\pr\sigma_{\textrm{\tiny e}}}+\frac\pr{\pr\sigma_{\textrm{\tiny h}}}\right\}\!\left\{\frac{\sigma_\h}{m_\h^*}-\frac{\sigma_\e}{m_\e^*}\right\}\!\frac{\langle\Phi|\frac1{|\rr_{\textrm{\tiny eh}}|}|\Phi\rangle}{\langle\Phi|\Phi\rangle}.
  $$
Real physical quantities should be invariant under the electron-hole exchange symmetry defined by the exchange of their coordinates, their masses and their electric charges
  \begin{equation}
\rr_{\e,\h}\rightarrow\rr_{\h,\e},~~m_{\e,\h}^*\rightarrow m_{\h,\e}^*,~~e\rightarrow-e. \label{symetrie}
  \end{equation}
As we expect that $\sigma_{\e,\h}\propto m_{\e,\h}^*eE_\dd R^2$, under an electron-hole exchange, the variational parameters should transform as $\sigma_{\e,\h}\rightarrow-\sigma_{\h,\e}$. Therefore, the trial function $\Phi(\rr_\ee,\rr_\h)$, the confinement Hamiltonian $H_0$, the Coulomb potential $V_\C(r_{\e\h})$ and the interaction Hamiltonian $W_\e(\rr_\e)+W_\h(\rr_\h)$ should remain invariant under the electron-hole exchange, as well as the norm of $\Phi(\rr_\e,\rr_\h)$ and the mean value of these operators. But, the further contribution will not, since it changes as $K(\sigma,\sigma_\e,\sigma_\h)\rightarrow-K(\sigma,\sigma_\e,\sigma_\h)$. Because of the mean value of the confinement Hamiltonian $H_0$ invariance, it should not bring any new contribution to real physical quantities, and should be discarded from $\ds\frac{\langle\Phi|H_0|\Phi\rangle}{\langle\Phi|\Phi\rangle}$, as given by Eq. \eref{Phi|H_0|Phi}.
\subsection{Stark effect in strong confinement regime}
As there seems to be no way to analytically compute the integrals in Eqs. \eref{Phi|Phi}, we propose to Taylor expand them in the strong confinement regime, {\it i.e.} when $\sigma R\ll1$. To perform this expansion, we have to specify the QD radii region, in which the following expressions are valid. For this, we shall assume that the quantities $\rho_{\e,\h}(\xi)\sigma R$ in the arguments of the functions $\exp(x)$ and $\sinh(x)$, appearing in Eq. \eref{Phi|Phi} should be sufficiently small. As $\rho_{\e,\h}(\xi)<\ds\frac32$ for all $\xi\in[-1,1]$, we will only consider the range of QD radii such that $\ds R\leq\frac2{3\sigma'}a^*$, so that $\rho_{\e,\h}(\xi)\sigma Rx,\rho_{\e,\h}(\xi)\sigma Ry\lesssim1$, if $-1\leq\xi\leq1$ and $0\leq x\leq y\leq1$. Thanks to the consistency condition Eq. \eref{consistency_condition}, the variational parameters $\sigma$ and $\sigma_{\e,\h}$ can be of the same order of magnitude. Then, we deduce, up to third order in $\sigma R$, that
  \begin{equation}
\fl\left\{
    \begin{array}{rcl}
\langle\Phi|\Phi\rangle\!\!\!\!\lnd=\rnd\!\!\!\!\ds1-B\sigma R+C\sigma^2R^2-D\sigma^3R^3
\\
\lnd\rnd\!\!\!\!\!\!\!\!\!\!\!\!\!\!\!\!\!\!\!\!\ds+(C-D'\sigma R)\frac{\sigma_\e^2+\sigma_\h^2}6R^2-D''
\sigma\sigma_\e\sigma_\h R^3+\Or(\sigma^4R^4),
\\
\ds\frac{\langle\Phi|V_\C(\rr_{\e\h})|\Phi\rangle}{\langle\Phi|\Phi\rangle}\!\!\!\!\lnd=
\rnd\!\!\!\!\ds\frac{-e^2}{\kappa R}\!\left\{A+B'\sigma R+C'\sigma^2R^2+C'_1(\sigma_\e^2+\sigma_\h^2)R^2\right.
\\
\lnd\rnd\!\!\!\!\!\!\!\!\!\!\!\!\!\!\!\!\!\!\!\!\left.+C'_2\sigma_\e\sigma_\h R^2+\Or(\sigma^3R^3)\right\}\!,
\\
\ds\frac{\langle\Phi|W_\e(\rr_\e)+W_\h(\rr_\h)|\Phi\rangle}{\langle\Phi|\Phi\rangle}\!\!\!\!\lnd=
\rnd\!\!\!\!\ds-eE_\dd(\sigma_\e+\sigma_\h)R^2\!\left\{\frac C3-C''\sigma R+\Or(\sigma^2R^2)\right\}\!.
    \end{array} \label{Phi|Phi_fort}
  \right.
  \end{equation}
The mean value of the total Hamiltonian $H$, under the influence of the electric field on both electron and hole is then expressed as an expansion in powers of the variational parameters, up to third order in $\sigma R$,
  \begin{eqnarray}
\frac{\langle\Phi|H|\Phi\rangle}{\langle\Phi|\Phi\rangle}\lnd=\rnd E_{\e\h}-A\frac{e^2}{\kappa R}+\frac{E^*}4\sigma'^2+\frac{\sigma_\e^2}{8m^*_\e}+\frac{\sigma_\h^2}{8m^*_\h}-2B'E^*\sigma' \nt
\\
\lnd\rnd~~~~-2C'E^*\frac R{a^*}\sigma'^2-C'_1\frac{\sigma_\e^2+\sigma_\h^2}\mu\frac R{a^*}-C'_2\frac{\sigma_\e\sigma_\h}\mu\frac R{a^*} \nt
\\
\lnd\rnd~~~~-eE_\dd(\sigma_\e+\sigma_\h)R^2\!\left\{\frac C3-C''\frac R{a^*}\sigma'\right\}\!+\cdots \label{Phi|H|Phi}
  \end{eqnarray}
We minimize the previous matrix element with respect to $\sigma'$ and $\sigma_{\e,\h}$ to obtain an approximate value of the ground state total energy. Their values are determined to the first order in $\ds\frac R{a^*}$ to insure the coherence of the expansion we made as
  \begin{equation}
\fl  \left\{
    \begin{array}{rcl}
\ds\sigma'_0\!\!\!\!\lnd=\rnd\!\!\!\!\ds4B'\!\left\{1+8C'\frac R{a^*}\right\}\!-\frac83CC''(m^*_\e+m^*_\h)\frac{e^2E_\dd^2R^4}{E^*}\frac R{a^*},
\vs{.2cm}
\\
\ds\sigma_{\e,\h}^0\!\!\!\!\lnd=\rnd\!\!\!\!\ds\frac{4C}3m_{\e,\h}^*eE_\dd R^2\!\left\{1+4\!\left[2C'_1\frac{m^*_{\e,\h}}\mu+C'_2\frac{m^*_{\h,\e}}{\mu}-\frac{3B'C''}C\right]\!\frac R{a^*}\right\}\!. \label{sigma'_0}
    \end{array}
  \right.
  \end{equation}
The Stark shift 
is identified as terms scaling as $\propto E_\dd^2$, and\footnote{From Eqs. \eref{Phi|Phi_fort}, we confirm that $K(\sigma,\sigma_\e,\sigma_\h)$, the contribution discarded from the confinement Hamiltonian $H_0$ mean value because of the electron-hole exchange symmetry, does not contribute to Stark effects
  $$
K(\sigma,\sigma_\e,\sigma_\h)=\!\left\{\frac{C'_1}2+\frac{C'_2}4\right\}\!\left\{\frac{\sigma_\h}{m_\h^*}-\frac{\sigma_\e}{m_\e^*}\right\}\!\left\{\sigma_\e+\sigma_\h\right\}\!\sigma R+\dots
  $$
Such terms contribute to the total Hamiltonian mean value $H$ up to the third order in $\ds\frac R{a^*}$. These third order terms just contribute to the variational parameters $\sigma'_0$ and $\sigma^0_{\e,\h}$ up to the first order, but not at all to the electron-hole pair ground state energy. Therefore, the last possible contribution to Stark effect should come from $K(\sigma'_0,\sigma^0_\e,\sigma^0_\h)$, where we replace the different variational parameters by their respective zeroth order expressions $\sigma'_0\approx4B'$ and $\sigma_{\e,\h}^0\approx\ds\frac{4C}3m_{\e,\h}^*eE_\dd R^2$, according to Eqs. \eref{sigma'_0}. Then, we straightforwardly verify that $K(\sigma'_0,\sigma^0_\e,\sigma^0_\h)=0$, which is what was expected.}
  \begin{equation}
\Delta E^{\textrm{\scriptsize strong}}_{\textrm{\scriptsize Stark}}=
-\Gamma_{\textrm{\scriptsize var}}(m^*_\e+m^*_\h) e^2E_\dd^2R^4\!\left\{1+8\Gamma^{\e\h}_{\textrm{\scriptsize var}}\frac R{a^*}\right\}\!, \label{DeltaE}
  \end{equation}
where $\Gamma_{\textrm{\scriptsize var}}$ appears as of universal character, while the constant $\Gamma^{\e\h}_{\textrm{\scriptsize var}}$ depends on the semiconductor. In terms of the effective masses of the electron and the hole, it can be expressed as
  $$
\Gamma^{\e\h}_{\textrm{\scriptsize var}}=C'_1\!\left\{\frac{m^*_\e}{m^*_\h}+\frac{m^*_\h}{m^*_\e}\right\}\!+C'_2-\frac{3B'C''}C.
  $$
The first contribution to this shift is simply the sum of the Stark shift contributions undergone by the ground states of both electron and hole taken individually as computed in Subsection {\it \ref{subsec_III_A}}. Because of the dependence of the constant $\Gamma^{\e\h}_{\textrm{\scriptsize var}}$ on the effective masses $m^*_{\e,\h}$, the second contribution to $\Delta E^{\textrm{\scriptsize strong}}_{\textrm{\scriptsize Stark}}$ indicates the existence of a further coupling between the electron and the hole, which appears as a standard dipolar interaction between two opposite electric charge carriers. This interpretation is a question of point view. Until now, we have considered that the interaction between the electron or the hole with the external electric field takes place individually, whereas they interact only through the Coulomb potential. This physical description justifies {\it a priori} the validity of the strong confinement regime assumption, for which the exciton states consist of uncorrelated individual confined electron and hole states. It then allows to intuitively build a coherent model in order to describe the QCSE in QDs in this regime, and also simplifies the calculations in practice. In spite of these advantageous properties, the previous remark suggests that this picture should be revised.

Actually, the total Hamiltonian electric field interaction part $W_\e(\rr_\e)+W_\h(\rr_\h)$ should also be written as 
  $$
W(\rr_{\e\h})=W_\e(\rr_\e)+W_\h(\rr_\h)=\E_\dd\cdot\ddd_{\e\h},
  $$
where $\ddd_{\e\h}=e\rr_{\e\h}$ is the exciton electric dipole moment. This is the standard dipolar interaction Hamiltonian of an electric dipole. It satisfies the electron-hole exchange symmetry, while the individual interaction Hamiltonian $W_{\e,\h}(\rr_{\e,\h})$ transform themselves one into another. In the strong confinement regime, despite the importance of confinement effects on excitonic ones, the dipolar interaction point of view expresses the remnant of electron-hole pair states, thought as exciton bound states under the influence of the electric field. It suggests the inclusion of a further term in the Hamiltonian $H$ describing the exciton-electric field interaction, which accounts for the polarization energy of the electron-hole pair, due to the difference between the dielectric constants of the semiconductor QD and the surrounding insulating matrix.
\subsection{Comparison with experimental data} \label{subsec_III_D}
In order to test the relevance of our model, we shall compare our predictions to real experimental data given in  \cite{Nomura_1990} and to other computational data of \cite{Nomura_1990b} in spherical $CdS_{0.12}Se_{0.88}$ micro-crystallites with material parameters: $\varepsilon=9.3$, $m_\e^*=0.13m_\e$, $m_\h^*=0.46m_\e$, $E^*=16$meV and $a^*=49$\AA, where $m_\e$ is the electron bare mass. The electric field amplitude inside the micro-crystal is fixed at $E_\dd=12.5$kV.cm$^{-1}$. We note that an earlier work \cite{Wen_1995} has used numerical diagonalization of the total Hamiltonian $H$, with the same material parameters for spherical $CdS_{0.12}Se_{0.88}$ QDs as well as electric field, to obtain theoretical predictions.
\subsubsection{Case of real experimental data}
  \begin{figure*}
\caption{Stark shift for confined interactive electron-hole pair as a function of the QD radius including the Coulomb interaction and excluding the polarization energy up to the zeroth (---) or to the first (--$\!~$--) order in comparison with experimental results (+) \cite{Nomura_1990}. $\Gamma^{\textrm{\tiny eh}}_{\textrm{\tiny var}}\approx-0.1629$.}
\label{figure_1}
    \begin{center}
\input{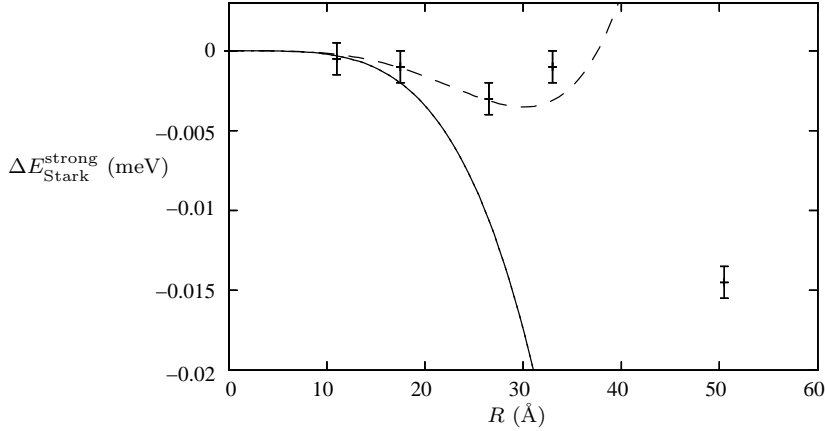}
    \end{center}
  \end{figure*}
\

\noindent \Fref{figure_1} presents a comparison between results we obtain and experimental values for spherical $CdS_{0.12}Se_{0.88}$ micro-crystallites \cite{Nomura_1990}. Two exciton peaks are experimentally resolved, which are attributed to the transitions from the highest valence sub-band and from the spin-orbit split-off state to the lowest conduction sub-band, with an energy splitting is about 0.39eV independently of the QD radius \cite{Nomura_1990}. The experimental values depicted by crosses in \Fref{figure_1} consist of mean values of the Stark shift of these two types of excitons. They seem to indicate that the Coulomb interaction is sufficient to explain correctly the amplitude of the Stark effects experimentally observed, as we expect, in the range of validity of QD radii.

In the strong confinement regime, our approach offers a model capable to describe QCSE at least for QD sizes $\ds \frac R{a^*}\leq\frac2{3\sigma'_0}$. But, $\sigma'_0$ is itself a function of $\ds\frac R{a^*}$, which is still considered as a small dimensionless parameter in the strong confinement regime. Here, the part of $\sigma_0'$ which depends on the electric field is negligible, because it scales as $\propto m_\e e^2E_\dd^2R^4$. This is at least of the same order of magnitude as the exciton Rydberg energy as soon as $R\leq50$\AA, if the electric field amplitude is fixed at $E_\dd=12.5$kV.cm$^{-1}$, and $\ds \frac{CC''(m^*_\e+m^*_\h)}{12B'C'}\approx0.0552m_\e$. Hence, according to Eqs. \eref{sigma'_0}, $\ds\sigma'_0\approx4B'\!\left\{1+8C'\frac R{a^*}\right\}\!.$ Therefore, up to first order in $\ds\frac R{a^*}$, our predictions should be valid for QD radii
  \begin{equation}
R\leq\frac{a^*}{2(3B'+4C')}\approx 0.6080a^*. \label{R}
  \end{equation}
According to this effective constraint, in the case of $CdS_{0.12}Se_{0.88}$ micro-crystals, this approach should lead to acceptable results in regard to experimental data as long as the cluster radius does not exceed 30\AA. \Fref{figure_1} shows that the absolute value of the Stark shift, computed up to the zeroth order, is significantly overestimated, except for a minor range of small QD radii compared with the one we expect. The results become much more accurate, if the Stark effects are computed up to the first order. In this case, \Fref{figure_1} exhibits a good agreement with the experimental data over the whole expected region of micro-crystals radii. In this domain of validity, the first order calculation seems to be efficient enough to describe QCSE in spherical semiconductor QDs. As soon as, the QD radius exceeds the maximal value for which the strong confinement regime is valid, our results diverge significantly from experimental data. Furthermore, Figure 4 in \cite{Wen_1995} shows Stark shifts of the ground  state and the two first excited states of the confined exciton, obtained by numerical diagonalization of the Hamiltonian $H$. Agreement with experimental data from \cite{Nomura_1990} is also reported in the ground state case. Whereas \cite{Wen_1995} gives a range 20\AA~$\leq R\leq100$\AA, in its validity domain $R\lesssim30$\AA, we may say that our analytical approach is totally consistent with the numerical approach of \cite{Wen_1995}.

To determine the maximal electric field amplitude, for which the weak electric field limit assumption remains valid, the consistency condition Eq. \eref{consistency_condition} is reconsidered for the electron or for the hole, in which the respective variational parameters are replaced by their variational values. After summing the expressions for the electron and for the hole, up to first order in $\ds\frac R{a^*}$, we deduce that
  $$
\frac{E_{\textrm{\scriptsize ele}}}{E_{\e\h}}\leq\frac1{\pi^2\ee C}\frac1{1+\frac43\frac{C'}{B'}}\approx0.1197.
  $$
Then, in the strong confinement regime, the hypothesis of weak electric field limit should be valid as soon as the typical electric dipole interaction energy does not represent more than about 12\% of the typical exciton confinement energy. If the micro-crystal radius is fixed at $R=10$\AA, the highest electric field amplitude for which the weak field limit assumption stays acceptable is about $E^{\textrm{\scriptsize max}}_\dd\approx450$kV.cm$^{-1}$. {\it Idem} if the QD radius is fixed near $R\approx30$\AA, the upper boundary of the strong confinement regime validity domain, the electric field amplitude inside the QD should not exceed $E^{\textrm{\scriptsize max}}_\dd\approx16.7$kV.cm$^{-1}$. These numerical results justify the choice of an electric field such that $E_\dd\approx12.5$kV.cm$^{-1}$ to compare theoretical predictions against experimental results, because it satisfies the weak field limit all along the strong confinement range of QD radius.

In a more general manner, as soon as the semiconductor of the synthesized QD is chosen, the strong confinement regime domain of validity and the weak electric field limit condition consist of a set of two constraints, which should be optimized by choosing the QD radius and the electric field amplitude as functions of the Bohr radius, the Rydberg energy and the confinement energy of a trapped exciton. 
But, for future technological applications, this set of constraints will permit to determine conversely the best possible semiconductor for practical and technological reasons, by imposing the typical QD size and the order of magnitude of the maximal electric field amplitude to use.
\subsubsection{Case of computational data} \label{subsubsec_III_D_2}
\

\noindent In the early nineties, a variational calculation on the same total Hamiltonian $H$ with computational is performed, in order to study Stark effect in spherical micro-crystals \cite{Nomura_1990b}. The weak field limit is also considered and the Hamiltonian mean value is expanded in powers of the variational parameters $\sigma_{\e,\h}$ up to the second order. However, terms scaling as $\propto\sigma_{\e,\h}^2$ are neglected, while those scaling as $\propto\sigma_\e\sigma_\h$ are kept. Our approach suggests that both terms have the same order of magnitude and contribute to the electron-hole pair Stark shift. Leaving out such contributions implies that $C'_1$ should vanish in Eq. \eref{Phi|H|Phi}. The expected Stark shift should be then affected, because $\ds\Gamma^{\e\h}_{\textrm{\scriptsize var}}=C'_2-\frac{3B'C''}C$. Then, $\Gamma^{\e\h}_{\textrm{\scriptsize var}}\approx-0.0333$ is independent of the semiconductor, and hardly represents about 20\% of its value, when we account for contributions scaling as $\propto\sigma_{\e,\h}^2$, {\it i.e.} if $C'_1\neq0$. Although the approximation made in \cite{Nomura_1990b} deeply changes the nature of Stark effect, and does not seem to describe correctly experimental results, except for very small QDs. \Tref{table_1} shows good agreement between our results if $C'_1=0$ and computational ones from \cite{Nomura_1990b}, even outside the validity domain of the strong confinement regime, {\it i.e.} for $R\geq30$\AA. This signifies that, in the strong confinement regime, a first order expansion in $\ds\frac R{a^*}$ of the ground state energy Stark shift should be sufficient, at least when only the Coulomb interaction is included in the electron-hole pair Hamiltonian, because there is no particular constraint on the variational parameter $\sigma$ in \cite{Nomura_1990b}.
\begin{table*}
\caption{Stark shift for confined interactive electron-hole pair as a function of the QD radius including the Coulomb interaction and excluding the polarization energy, where terms scaling as $\propto\sigma_{\e,\h}^2$ are removed from the total Hamiltonian $H$ mean value, {\it i.e.} $C'_1=0$ and $\Gamma^{\textrm{\tiny eh}}_{\textrm{\tiny var}}\approx-0.0333$, in comparison with computational results \cite{Nomura_1990b} in spherical $CdS_{0.12}Se_{0.88}$ micro-crystallites.} \label{table_1}
    \begin{center}
{\footnotesize
      \begin{tabular}{ccccccccc}
\br
R & \AA & 10 & 20 & 30 & 40 & 50
\\
\hline
$-\Delta E_{\textrm{\tiny num}}$ & meV & 2.08 10$^{-4}$ & 3.16 10$^{-3}$  & 1.49 10$^{-2}$  & 4.34 10$^{-2}$ & 9.63 10$^{-2}$
\\
$-\Delta E^{\textrm{\tiny strong}}_{\textrm{\tiny Stark}}$ & meV & 2.03 10$^{-4}$ & 3.07 10$^{-3}$  & 1.46 10$^{-2}$  & 4.31 10$^{-2}$ & 9.78 10$^{-2}$
\\
relative error &  & $<$3\% & $<$3\% & $\approx$2\% & $<$1\% & $<$2\%
\\
\br
      \end{tabular}}
    \end{center}
\end{table*}
\section{Stark effects with polarization energy} \label{sec_IV}
To investigate in more details Stark effects in semiconductor micro-crystallites and to especially integrate the electric dipole interaction point of view, we shall introduce the following polarization energy introduced in \cite{Brus_1984} to the total electron-hole Hamiltonian $H$
  $$
P(\rr_\e,\rr_\h)=\frac{e^2}{2R}\sum_{l\geq1}\alpha_l(\varepsilon_\rrr)\!\left\{\!\left(\frac{r_\e}R\right)^{\!\!2l}\!\!+\!\left(\frac{r_\h}R\right)^{\!\!2l}\!\!-2\!\left(\frac{r_\e r_\h}{R^2}\right)^{\!\!l}\PP_l(\cos\theta_{\e\h})\right\}\!,
  $$
where $\PP_l(x)$ denotes the Legendre polynomial of index $l$ and of variable $x$, and where we define, for a later purpose, the constants $\alpha_l(\varepsilon_\rrr)$ as functions of the relative dielectric constant $\varepsilon_\rrr$ as $\ds\alpha_l(\varepsilon_\rrr)=\frac{(l-1)(\varepsilon_\rrr-1)}{\kappa(l\varepsilon_\rrr+l+1)},~l\in\mathbb N^*$. We wish to apply a variational method to the new Hamiltonian $H'=H+P$, which takes into account the polarization energy $P(\rr_\e,\rr_\h)$. For this, we keep the form of the variational trial function $\Phi(\rr_\e,\rr_\h)$, since the polarization energy should not basically change the properties of the trial function.
\subsection{Strong confinement regime}
Here, the polarization energy mean value is expressed as an expansion in the variational parameters, which has the same form as the Coulomb potential mean value, where the constants $A$, $B'$, $C'$, $C'_1$ and $C'_2$ are replaced by functions of the relative dielectric constant $\varepsilon_\rrr$. This is the reason why we adopt the same notations for these quantities, but with explicit dependence on $\varepsilon_\rrr$, as shown in \Tref{table_5}. In fact, the polarization energy mean value in the quantum state defined by $\Phi(\rr_\e,\rr_\h)$ should be written as
  \begin{eqnarray}
\fl\frac{\langle\Phi|P(\rr_\e,\rr_\h)|\Phi\rangle}{\langle\Phi|\Phi\rangle}\lnd=\rnd\ds-\frac{e^2}{\kappa R}\!\left\{A(\varepsilon_\rrr)+B'(\varepsilon_\rrr)\sigma R+C'(\varepsilon_\rrr)\sigma^2R^2\right. \nt
\\
\fl\lnd\rnd~~~~\left.+C'_1(\varepsilon_\rrr)(\sigma_\e^2+\sigma_\h^2)R^2+C'_2(\varepsilon_\rrr)\sigma_\e\sigma_\h R^2+\Or(\sigma^3R^3)\right\}\!. \label{Phi|P(rr_e,rr_h)|Phi}
  \end{eqnarray}
In this formalism, we obtain expressions for the variational parameters and the Stark shift considering only the polarization energy or both the Coulomb interaction and the polarization energy in Stark effects by Eqs. \eref{sigma'_0} and \eref{DeltaE} in Section {\bf \ref{sec_III}}, by replacing all the appearing constants, in the first case, by the corresponding functions of $\varepsilon_\rrr$ and, in the second case, by the sum of both contributions. An important step in this calculation consists of the simple idea of rewriting the functions $\alpha_l(\varepsilon_\rrr)$ in a such way that it becomes possible to perform analytically the summation of the series appearing in the expression of the polarization energy $P(\rr_\e,\rr_\h)$. As matter of fact, we remark that for $l\in\mathbb N^*$
  $$
\alpha_l(\varepsilon_\rrr)\approx\frac{\varepsilon_\rrr-1}{\kappa(\varepsilon_\rrr+1)}\!\left\{1+\frac{\varepsilon_\rrr}{(\varepsilon_\rrr+1)l}\right\}\!=\widetilde\alpha_l(\varepsilon_\rrr).
  $$
Introducing $\Delta_l(\varepsilon_\rrr)$ for $l\in\mathbb N^*$ as the relative error between the functions $\alpha_l(\varepsilon_\rrr)$ and $\widetilde\alpha_l(\varepsilon_\rrr)$, the replacement of $\alpha_l(\varepsilon_\rrr)$ by $\widetilde\alpha_l(\varepsilon_\rrr)$ is reasonable because it leads to negligible relative errors: {\it e.g.} for $CdS_{0.12}Se_{0.88}$ micro-crystallites, for which $\varepsilon_\rrr=4.0$, we get that $\Delta_1(\varepsilon_\rrr)\approx8\%$, $\Delta_2(\varepsilon_\rrr)\approx3\%$ and $\ds\Delta_l(\varepsilon_\rrr)=\frac{\varepsilon_\rrr}{(\varepsilon_\rrr+1)^2}\frac1{l(l+1)}\lesssim1\%$, for $l\geq3$. The polarization energy is then written as
  \begin{eqnarray}
\fl P(\rr_\e,\rr_\h)\lnd=\rnd\frac{e^2}{2\kappa R}\frac{\varepsilon_\rrr-1}{\varepsilon_\rrr+1}\!\left[\frac1{1-\frac{r^2_\e}{R^2}}+\frac1{1-\frac{r^2_\h}{R^2}}-\frac{\varepsilon_\rrr}{\varepsilon_\rrr+1}\!\left\{\log\!\left(1-\frac{r_\e^2}{R^2}\right)\!+\log\!\left(1-\frac{r_\h^2}{R^2}\right)\!\right\}\right. \nt
\\
\fl\lnd\rnd~~~~\left.-2\sum_{l\geq0}\left(\frac{r_\e r_\h}{R^2}\right)^{\!\!l}\PP_l(\cos\theta_{\e\h})-\frac{2\varepsilon_\rrr}{\varepsilon_\rrr+1}\sum_{l\geq1}\frac1l\!\left(\frac{r_\e r_\h}{R^2}\right)^{\!\!l}\PP_l(\cos\theta_{\e\h})\right]\!, \label{P}
  \end{eqnarray}
where it is possible to sum {\it a priori} the series in the polarization energy because of the confining potential well. The mean value of $P(\rr_\e,\rr_\h)$ in the quantum state $\Phi(\rr_\e,\rr_\h)$ contains only contributions of electron and hole such that $r_{\e,\h}<R$.
\subsection{Comparison with experimental data}
The experimental parameters are those of Subsection \ref{subsec_III_D} and the electric field amplitude inside the micro-crystal is kept at $E_\dd=12.5$kV.cm$^{-1}$.
\subsubsection{Case of real experimental data}
\

\noindent We begin to evaluate the strong confinement regime validity region by considering Eq. \eref{R}, on which we apply the appropriate changes according to the cases. If we only consider the polarization energy, the strong confinement regime is not actually valid, because it turns out that $\sigma'_0\leq0$, in this case. Therefore, the interaction part of the variational function should be $\phi_{\textrm{\scriptsize rel}}(r_{\e,\h})=\ee^{|\sigma'_0|r_{\e,\h}}$. This reveals that the polarization energy is repulsive, while our approach is build on an attractive point of view for the interaction between the electron and the hole. If we consider both Coulomb potential and polarization energy, this problem no longer exists because the attractive effects of Coulomb potential are more important than the repulsive ones due to the polarization energy, so that $\sigma'_0$ remains positive. Thus, the strong confinement regime remains valid up to $R\lesssim 1.0179a^*$, {\it i.e.} $R\lesssim50$\AA~for $CdS_{0.12}Se_{0.88}$ micro-crystallites. This domain of validity is twice as large as the one obtained if only the Coulomb potential is taken into account. In the same spirit, we evaluate an order of magnitude for the maximal electric field amplitude, for which the weak field limit should be assumed to be valid, as $\ds\frac{E_{\textrm{\scriptsize ele}}}{E_{\e\h}}\lesssim0.1186$, when only polarization is taken into account, and $\ds\frac{E_{\textrm{\scriptsize ele}}}{E_{\e\h}}\lesssim0.1204$, when polarization energy and Coulomb interaction are both included. From such numerical values, we should remark that the weak field limit is relatively insensitive to the interaction between the charge carriers. It may notably signify that this limit is relevantly chosen because it is an independent condition coming from the strong confinement regime validity domain. For example, this corresponds, for a QD radius of 30\AA, to an electric field amplitude of about $E^{\textrm{\scriptsize max}}_\dd=16.5$kV.cm$^{-1}$, on one hand, and to an electric field amplitude of about $E^{\textrm{\scriptsize max}}_\dd=16.8$kV.cm$^{-1}$, on the other hand.

  \begin{figure*}
\caption{Stark shift for confined interactive electron-hole pair as a function of the QD radius including both the Coulomb interaction and the polarization energy up to the zeroth (---) or to the first order (--$\!~$--), where $\Gamma^{\textrm{\tiny eh}}_{\textrm{\tiny var}}\approx-0.2045$, and including only the polarization energy up to the first order (--$\!~\cdot\!~$--), where $\Gamma^{\textrm{\tiny eh}}_{\textrm{\tiny var}}\approx-0.0416$, in comparison with experimental results (+) \cite{Nomura_1990} in spherical $CdS_{0.12}Se_{0.88}$ micro-crystallites.} \label{figure_2}
    \begin{center}
\input{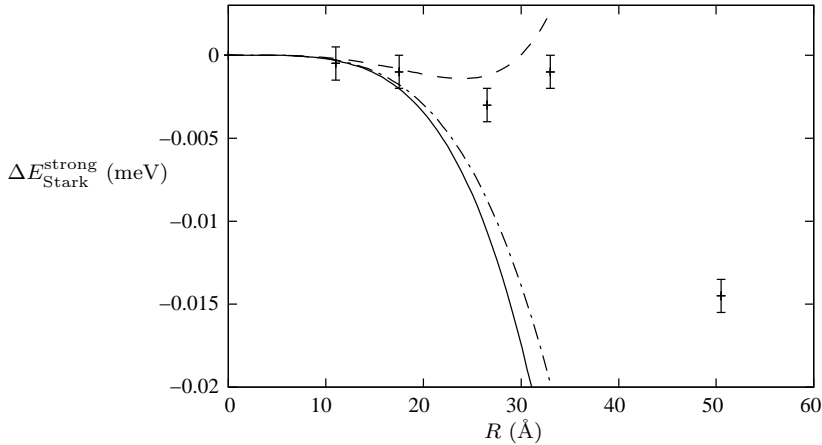}
    \end{center}
  \end{figure*}
As the electric field amplitude is fixed inside the $CdS_{0.12}Se_{0.88}$ micro-crystallites at $E_\dd=12.5$kV.cm$^{-1}$, the weak electric field limit should be satisfied in the range of QD radii $R\leq30$\AA~and we can compare, at least in this domain, our predictions to experimental data. In the rest of the strong confinement regime validity domain, {\it i.e.} for radii 30\AA~$\lesssim R\lesssim50$\AA, the weak field limit is no longer valid. This may actually explain the significant divergence from experimental results in this region. Once again the analogy with rectangular quantum boxes should be quite helpful. In the opposite limit of strong electric field, the Stark shift undergone by the ground state of the electron-hole pair confined in a rectangular QD shows essentially a linear behavior in the electric field amplitude \cite{Wei_2007}, which is a totally different behavior in comparison to the quadratic dependence found here. \Fref{figure_2} clearly shows that the behavior of the polarization energy, if considered alone, is not satisfactory because it seems that its contribution does not counterbalance the divergence of the zeroth order Stark shift for QD radii near the upper boundary of the strong confinement validity region, while \Fref{figure_1} suggests that the contribution to the Stark effects due to the Coulomb potential seems to consist of more important contributions. \Fref{figure_2} stresses this point of view. In the strong confinement regime, if we account for the Coulomb potential and the polarization energy, the results accurately fit the experimental data, except if the QD size begins to reach the lower boundary of the domain of this regime, in which the weak field limit is not valid anymore. The reason for this phenomenon is simple to understand. For such QD sizes, we remark that the first order term has the same order of magnitude as the zeroth order contribution to the Stark shift. Then, if it is reasonable to maintain  {\it a priori} a perturbation point of view, the first order correction is not sufficient to describe correctly QCSE in the whole domain of strong confinement regime validity. It is perhaps advisable to continue the expansion to one or two further orders and to revise the definition of weak field limit. However, the computations become quite involved and we think that such approach does not really bring a significant improvement to the understanding of the Stark effects in QDs.
\subsubsection{Case of computational data}
\

\noindent Following the reasonings made in Subsubsection {\it \ref{subsubsec_III_D_2}}, we add the polarization energy $P(\rr_\e,\rr_\h)$ to the total Hamiltonian of the electron-hole pair but impose that $C'_1(\varepsilon_\rrr),C'_1=0$, in such a way that terms scaling as $\propto\sigma_{\e,\h}^2$ do not contribute to the electron-hole ground state Stark shift. Once again, the new value for $\Gamma^{\e\h}_{\textrm{\scriptsize var}}$ represents a small part from the one, in which all contributions are kept. Likewise, when we only consider the Coulomb interaction, \Tref{table_2} shows that there is still a good agreement between our results and computational ones from \cite{Nomura_1990b}, over the whole validity domain of strong confinement regime. The divergence from experimental results is still significant after inclusion of polarization energy, even for small QD radius. This confirms that the terms scaling as $\propto\sigma_{\e,\h}^2$ play a relevant role in Stark effects and should not be discarded. While, this also legitimates the approximation $\alpha_l(\varepsilon_\rrr)\approx\widetilde\alpha_l(\varepsilon_\rrr)$, because \Tref{table_2} suggests that this approximation does not seem over-estimate the polarization energy contribution to the Stark shift. Maybe, each term of the sum defining the polarization energy plays a role in Stark effect, but errors made term by term should not cumulate. A first order expansion in $\ds\frac R{a^*}$ of the ground state energy Stark shift should not be sufficient and reinforces the idea according which it is necessary to carry on the expansion at least up to the second order. If the second order expansion does not improve the situation, this signifies that there should exist another reason for this divergence. Therefore, because the strong confinement regime validity domain is not affected be dropping of terms scaling $\propto\sigma_{\e,\h}^2$, it is reasonable to think that it comes from the failure of the weak field limit assumption for QD radius sufficiently close to its upper boundary.
\begin{table*}
\caption{Stark shift for confined interactive electron-hole pair as a function of the QD radius including the Coulomb interaction and the polarization energy where terms scaling as $\propto\sigma_{\e,\h}^2$ are suppressed from the total Hamiltonian $H$ mean value, {\it i.e.} $C'_1(\varepsilon_\rrr),C'_1=0$ and $\Gamma^{\textrm{\tiny eh}}_{\textrm{\tiny var}}\approx-0.0446$, in comparison with computational results \cite{Nomura_1990b} in spherical $CdS_{0.12}Se_{0.88}$ micro-crystallites.} \label{table_2}
    \begin{center}
{\footnotesize
      \begin{tabular}{ccccccccc}
\br
R & \AA & 10 & 20 & 30 & 40 & 50
\\
\hline
$-\Delta E_{\textrm{\tiny num}}$ & meV & 2.09 10$^{-4}$ & 3.14 10$^{-3}$  & 1.45 10$^{-2}$  & 4.11 10$^{-2}$ & 8.93 10$^{-2}$
\\
$-\Delta E^{\textrm{\tiny strong}}_{\textrm{\scriptsize Stark}}$ & meV & 1.99 10$^{-4}$ & 2.94 10$^{-3}$  & 1.36 10$^{-2}$  & 3.90 10$^{-2}$ & 8.55 10$^{-2}$
\\
relative error &  & $\approx$5\% & $\approx$6\% & $\approx$6\% & $\approx$5\% & $\approx$4\%
\\
\br
      \end{tabular}}
    \end{center}
\end{table*}
\section{Conclusion}
By considering a simple EMA model under the assumptions of strong confinement regime by a infinite potential well and of weak electric field limit, we are able to obtain analytical results for Stark effect in semiconducting micro-crystallites with spherical shape. In the domain of validity of physical approximations, the numerical values we can deduce agree with experimental data. We furthermore clarify why other variational calculations predict numerical results, which markedly diverge from experimental ones.

Despite these successes, our approach has been invalidated in a particular range of QD sizes, for which the strong confinement point of view still holds, but for which the weak electric field limit assumption fails. Thus, a future research work may focus on trying to apply the strong confinement regime in the limit of strong electric field indeed, even in a more general manner, to any electric field amplitude. The case of the weak confinement regime of the electron-hole pair is much more difficult, even in the weak field limit. Actually, in such a regime, the integration domain of integrals, we have to deal with, to compute the square of the trial function norm or the diagonal matrix elements of physical operators consists of a half-rectangle instead of the domain $\mathcal D$, which is a half-square. This implies an explicit break down of the electron-hole exchange symmetry. Then, a different approach may be needed.
\appendix
\section{Constants} \label{appendix_B}
In the following tables, we sum up all appearing constants and give their approximate values. The function $\ds\Si(x)=\int_0^x\frac{\dd t}t\sin(t)$ denotes the standard sine integral.
\subsection{Constants occurring in Stark effect expressions in absence of polarization}
\Tref{table_3} presents analytical expressions and approximate values of constants when only the Coulomb potential is taken into account.
\begin{table*}
\caption{Definition, analytical expression and approximate value of constants when only the Coulomb interaction is taken into account.}
\label{table_3}
    \begin{center}
{\footnotesize
      \begin{tabular}{ccc}
\br
Name & Expression & Value
\\
\hline
$A$ & $\ds2-\frac1\pi\!\left\{\Si(2\pi)-\frac{\Si(4\pi)}2\right\}$ & 1.7861
\vs{.1cm}
\\
$B_1$ & $\ds\frac23-\frac5{8\pi^2}$ & 0.6033
\vs{.1cm}
\\
$B_2$ & $\ds\frac29+\frac{13}{24\pi^2}+\frac1{2\pi^3}\!\left\{\Si(2\pi)-\frac{\Si(4\pi)}2\right\}$ & 0.2879
\vs{.1cm}
\\
$B$ & $\ds B_1+\frac{B_2}3$ & 0.6993
\\
$B'$ & $AB-1$ & 0.2489
\\
$C$ & $\ds\frac13-\frac1{2\pi^2}$ & 0.2827
\vs{.1cm}
\\
$C'$ & $\ds A(B^2-C)-\frac B2$ & 0.0189
\vs{.1cm}
\\
$C'_1$ & $\ds\frac{B_1-2AC}{12}$ & -0.0339
\vs{.1cm}
\\
$C'_2$ & $\ds\frac{B_2}{18}$ & 0.0160
\vs{.1cm}
\\
$D_1$ & $\ds\frac25-\frac{13}{8\pi^2}+\frac{147}{64\pi^4}$ & 0.2589
\vs{.1cm}
\\
$D_2$ & $\ds\frac2{15}-\frac1{8\pi^2}-\frac{21}{64\pi^4}$ & 0.1173
\vs{.1cm}
\\
$D_3$ & $\ds\frac2{25}+\frac{37}{120\pi^2}-\frac{1153}{320\pi^4}-\frac3{2\pi^5}\!\left\{\Si(2\pi)-\frac{\Si(4\pi)}2\right\}$ & 0.0710
\vs{.1cm}
\\
$D$ & $\ds\frac{5D_1+10D_2+D_3}{30}$ & 0.2539
\vs{.1cm}
\\
$D'$ & $\ds\frac{3D_1+4D_2+D_3}6$ & 0.2195
\vs{.1cm}
\\
$D''$ & $\ds\frac{5D_2-D_3}{45}$ & 0.0115
\vs{.1cm}
\\
$C''$ & $\ds\frac{D'+3D''-BC}3$ & 0.0187
\\
\br
      \end{tabular}}
    \end{center}
\end{table*}
\subsection{Constants occurring in Stark effect expressions in presence of polarization}
We evaluate the polarization energy mean value using Eq. \eref{P}. To this end, we compute integral representations of the polarization energy terms depending only on the radial coordinates $r_{\e,\h}$ of the electron and the hole following the reasoning, made in Subsection {\it \ref{subsec_III_B}} to get Eq. \eref{Phi|Phi}. This reasoning does not apply to the angular part. However, we are able to provide exact expressions for all the constants, which appears in the calculations, except for $\delta'''$, $\gamma'''$ and $\gamma''''$. For these, we obtain integral representations, which cannot be analytically computed at the moment. Their approximative values are computed numerically by using Wolfram Research Mathematica$^\textrm{\textregistered}$ 7. As exact expressions for other constants are quite cumbersome, we give only their approximate values.

Let us define the constants $\beta'$, $\beta''$, $\gamma'$, $\gamma''$, $\gamma'''$, $\gamma''''$, $\delta'$, $\delta''$ and $\delta'''$ by the expressions
  \begin{eqnarray*}
\lnd\rnd\langle\Phi|\frac1{1-\frac{r^2_\e}{R^2}}+\frac1{1-\frac{r^2_\h}{R^2}}|\Phi\rangle
\\
\lnd=\rnd\ds\beta'-\gamma'\sigma R+\delta'\sigma^2R^2+\frac{\delta'}6(\sigma_\e^2+\sigma_\h^2)R^2+\Or(\sigma^3R^3),
\\
\lnd\rnd-\langle\Phi|\log\!\left(1-\frac{r_\e^2}{R^2}\right)\!+\log\!\left(1-\frac{r_\h^2}{R^2}\right)\!|\Phi\rangle
\\
\lnd=\rnd\ds\beta''-\gamma''\sigma R+\delta''\sigma^2R^2+\frac{\delta''}6(\sigma_\e^2+\sigma_\h^2)R^2+\Or(\sigma^3R^3),
\\
\lnd\rnd2\langle\Phi|\sum_{l\geq0}\left(\frac{r_\e r_\h}{R^2}\right)^{\!\!l}\PP_l(\cos\theta_{\e\h})|\Phi\rangle
\\
\lnd=\rnd\ds2-\gamma'''\sigma R-\delta'''\sigma^2R^2+2C\!\left\{\sigma^2+\frac{\sigma_\e^2+\sigma_\h^2}6\right\}\!R^2-\frac{\sigma_\e\sigma_\h}{18}R^2+\Or(\sigma^3R^3),
\\
\lnd\rnd2\langle\Phi|\sum_{l\geq1}\frac1l\!\left(\frac{r_\e r_\h}{R^2}\right)^{\!\!l}\PP_l(\cos\theta_{\e\h})|\Phi\rangle
\\
\lnd=\rnd\ds-\gamma''''\sigma R-\delta'''\sigma^2R^2-\frac{\sigma_\e\sigma_\h}{18}R^2+\Or(\sigma^3R^3).
  \end{eqnarray*}
\Tref{table_4} presents approximate values for constants which appear in the polarization energy diagonal matrix element $\langle\Phi|P(\rr_\e,\rr_\h)|\Phi\rangle$, while \Tref{table_5} defines constants which appear in the polarization mean value Eq. \eref{Phi|P(rr_e,rr_h)|Phi} and gives their approximate values in $CdS_{0.12}Se_{0.88}$ micro-crystals.
\begin{table*}
\caption{Approximate value of constants appearing in $\langle\Phi|P(\rr_\e,\rr_\h)|\Phi\rangle$.} \label{table_4}
    \begin{center}
{\footnotesize
      \begin{tabular}{cccccccccc}
\br
Name & Value & Name & Value & Name & Value & Name & Value & Name & Value
\\
\hline
$\beta'$ & 3.1144 & $\gamma'$ & 2.3218 & $\gamma'''$ & 1.3263 & $\delta'$ & 0.9973 & $\delta'''$ & 0.0533
\\
$\beta''$ & 0.7524 & $\gamma''$ & 0.5992 & $\gamma''''$ & -0.0704 & $\delta''$ & 0.2708
\\
\br
      \end{tabular}}
    \end{center}
\end{table*}
\begin{table*}
\caption{Definition and approximate value of constants appearing in the polarization mean value Eq. \eref{Phi|P(rr_e,rr_h)|Phi} in $CdS_{0.12}Se_{0.88}$ micro-crystals.}\label{table_5}
    \begin{center}
{\footnotesize
      \begin{tabular}{ccc}
\br
Name & Expression & Value for $CdS_{0.12}Se_{0.88}$
\\
\hline
$\beta(\varepsilon_\rrr)$ & $\ds\frac12\frac{\varepsilon_\rrr-1}{\varepsilon_\rrr+1}\!\left\{\beta'-2+\frac{\varepsilon_\rrr}{\varepsilon_\rrr+1}\beta''\right\}\!$ & 0.5149
\vs{.1cm}
\\
$\gamma(\varepsilon_\rrr)$ & $\ds\frac12\frac{\varepsilon_\rrr-1}{\varepsilon_\rrr+1}\!\left\{\gamma'-\gamma'''+\frac{\varepsilon_\rrr}{\varepsilon_\rrr+1}(\gamma''-\gamma'''')\right\}\!$ & 0.4594
\vs{.1cm}
\\
$\delta_1(\varepsilon_\rrr)$ & $\ds\frac{\varepsilon_\rrr-1}{\varepsilon_\rrr+1}\!\left\{\delta'-2C+\frac{\varepsilon_\rrr}{\varepsilon_\rrr+1}\delta''\right\}\!$ & 0.3891
\vs{.1cm}
\\
$\delta_2(\varepsilon_\rrr)$ & $\ds\frac{\varepsilon_\rrr-1}2\frac{2\varepsilon_\rrr+1}{(\varepsilon_\rrr+1)^2}$ & 0.5400
\vs{.1cm}
\\
$\delta(\varepsilon_\rrr)$ & $\ds\frac{\varepsilon_\rrr-1}{\varepsilon_\rrr+1}\!\left\{\delta'+\delta'''-2C+\frac{\varepsilon_\rrr}{\varepsilon_\rrr+1}(\delta''+\delta''')\right\}\!$ & 0.4467
\vs{.1cm}
\\
$A(\varepsilon_\rrr)$ & $-\beta(\varepsilon_\rrr)$ & -0.4467
\vs{.1cm}
\\
$B'(\varepsilon_\rrr)$ & $-\beta(\varepsilon_\rrr)B+\gamma(\varepsilon_\rrr)$ & -0.0993
\vs{.1cm}
\\
$C'(\varepsilon_\rrr)$ & $\ds-\beta(\varepsilon_\rrr)(B^2-C)+\gamma(\varepsilon_\rrr)B-\frac{\delta(\varepsilon_\rrr)}2$ & -0.0083
\vs{.1cm}
\\
$C'_1(\varepsilon_\rrr)$ & $\ds-\frac{\delta_1(\varepsilon_\rrr)-2\beta(\varepsilon_\rrr)C}{12}$ & -0.0082
\vs{.1cm}
\\
$C'_2(\varepsilon_\rrr)$ & $\ds-\frac{\delta_2(\varepsilon_\rrr)}{18}$ & -0.0300
\\
\br
      \end{tabular}}
    \end{center}
\end{table*}

\end{document}